\documentclass[authoryear]{elsarticle}
\usepackage{latexsym}
\usepackage{float}
\usepackage[latin2]{inputenc}
\usepackage{graphicx}
\usepackage{ae}
\usepackage{aecompl}
\usepackage{setspace}
\usepackage{amsmath}
\usepackage{amssymb}
\usepackage[hidelinks]{hyperref}

\makeatletter
\def\ps@pprintTitle{%
	\let\@oddhead\@empty
	\let\@evenhead\@empty
	\def\@oddfoot{\centerline{\thepage}}%
	\let\@evenfoot\@oddfoot}
\makeatother

\begin{document}

\title{Introducing and Applying Newtonian Blurring: An Augmented Dataset of 126,000 Human Connectomes at {\tt braingraph.org}}
	
\author[p]{László Keresztes\corref{cor2}}
\ead{keresztes@pitgroup.org}
\author[p]{Evelin Szögi\corref{cor2}}
\ead{szogi@pitgroup.org}
\author[p]{Bálint Varga\corref{cor2}}
\ead{balorkany@pitgroup.org}
\author[p,u]{Vince Grolmusz\corref{cor1}}
\ead{grolmusz@pitgroup.org}
\cortext[cor1]{Corresponding author}
\cortext[cor2]{Joint first authors}
\address[p]{PIT Bioinformatics Group, Eötvös University, H-1117 Budapest, Hungary}
\address[u]{Uratim Ltd., H-1118 Budapest, Hungary}

\date{}

\begin{abstract}
	Gaussian blurring is a well-established method for image data augmentation: it may generate a large set of images from a small set of pictures for training and testing purposes for Artificial Intelligence (AI) applications. When we apply AI for non-imagelike biological data, hardly any related method exists. Here we introduce the ``Newtonian blurring'' in human braingraph (or connectome) augmentation: Started from a dataset of 1053 subjects, we first repeat a probabilistic weighted braingraph construction algorithm 10 times for describing the connections of distinct cerebral areas, then take 7 repetitions in every possible way, delete the lower and upper extremes, and average the remaining 7-2=5 edge-weights for the data of each subject. This way we augment the 1053 graph-set to 120 x 1053 = 126,360 graphs. In augmentation techniques, it is an important requirement that no artificial additions should be introduced into the dataset. Gaussian blurring and also this Newtonian blurring satisfy this goal. The resulting dataset of 126,360 graphs, each in 5 resolutions (i.e., 631,800 graphs in total), is freely available at the site \url{https://braingraph.org/cms/download-pit-group-connectomes/}. Augmenting with Newtonian blurring may also be applicable in other non-image related fields, where probabilistic processing and data averaging are implemented. 
\end{abstract}
	
\maketitle

\bigskip

\section*{Introduction} 

Gaussian blur operations \citep{Mair1996,Cho1991,Erdem1990} are standard tools in image processing for different purposes. In Artificial Intelligence (AI), it is widely used for image-data augmentation, i.e., increasing the size of the training and testing image databases by blurring some parts of the original images and adding the resulting pictures to the dataset \citep{Talukdar2018,Andriyanov2020,Lands2020}. The augmented dataset is applicable for pattern recognition tasks with AI tools, and -- since no new objects were artificially introduced, just some parts of the images blurred -- the integrity and the validity of the image-set will not be hurt. Additionally, in many applications in pattern recognition, the positive or negative identification of patterns is desirable even when some noise (i.e., blurring) is present in the images. 

In non-image datasets, and especially biological datasets, similar augmenting techniques are generally missing. In the present contribution, we describe an augmenting method, which is called ``Newtonian blurring'', paying tribute to Isaac Newton's Binomial Theorem and binomial coefficients \citep{Newton1687}. The Newtonian blurring is introduced and first applied here for augmenting a set of human braingraphs.

\subsection*{Human braingraphs}

Human braingraphs (or connectomes) describe the cerebral connections between the anatomically identified areas of the human brain. Human brain, in a simplified setting, consists of gray matter and white matter. Gray matter is -- again, with simplifications -- formed from the bodies (or somas) of the neurons, and the white matter from the axonal fibers of the neurons: these fibers connect the somas of the neurons. The white matter also contains myelin covers of these fibers, providing electrical insulation of the axonal fibers. The gray matter is situated on the outer surface (the cortex) of the brain, and also in some sub-cortical regions; the white matter can be found under the cortex, in the internal parts of the brain. 

It is a very interesting question to map the neuronal level connections of the brain: in this mapping, a graph is defined, where the nodes correspond to neurons, and an edge connects two neurons if the axon of one of them is connected to the other's dendrite. Unfortunately, no such graph can be measured for the human brain, which has 80 billion neurons: we simply do not have methods for discovering all the connections of such a great number of vertices. The nematode worm {\em Caenorhabditis elegans} is the only developed organism, for which such a graph is described \citep{White1986}, but it has only 203 neurons. Very recently, it is announced that the neuronal level connectome of the fruit fly {\em Drosophila melanogaster}  \citep{Zheng2018} is within our reach in one or two years, but today only the connections between some 25,000 neurons are available from the total of 100,000 neurons of the flybrain, i.e., the brain of the fruit fly \citep{Xu2020}.

The connectome (or the braingraph) of humans can be described today in a coarse resolution: the vertices correspond to around 1000 anatomically identified areas (called ROIs, Regions of Interests) of the gray matter. We write here ``areas'' instead of ``volumes'', since, in the cortex, the gray matter is relatively thin, typically 1-3 mm thick. The edges of the graph connect those pairs of vertices, between which a magnetic resonance imaging (MRI) workflow discovers axonal fibers \citep{Hagmann2008,Cammoun2012}. More exactly, diffusion MRI is capable of identifying the directional distribution of the water molecules in the neurons of the brain. This distribution is isotropic in the large body (soma) of the neurons, but it has a definitive direction (i.e., it is anisotropic) in the thin and long, tunnel-like axon of the neuron.  Therefore, axonal fibers can be discovered in the white matter of the brain, and one can construct braingraphs (or connectomes) as follows: two vertices, corresponding to two ROIs, are connected by an edge if at least one axonal fiber is discovered between them. The number of fibers detected is assigned as weights to the edges. For more details, we refer to \citep{Hagmann2008,Kerepesi2016b,Szalkai2016d,Varga2020}. 
	
As a result of a large, NIH-funded research project, the Human Connectome Project (HCP) \citep{McNab2013},  high-quality diffusion MRI data were published from hundreds of healthy, young adults. Applying this resource, together with an integrated toolset \citep{Daducci2012} for computing braingraphs from the MRI data, our research group successfully introduced numerous mathematical and graph-theoretical techniques into the analysis of the human braingraphs \citep{Szalkai2015a,Szalkai2015,Kerepesi2015b,Kerepesi2016b,Szalkai2016,Szalkai2016a,Kerepesi2015a,Kerepesi2016,Szalkai2016d,Szalkai2016e,Fellner2017,Fellner2018,Fellner2019,Fellner2019a}.

We have also published the braingraphs, computed by us at the site \url{https://braingraph.org}. First, we had made available several graph-sets, based on the HCP 500 Subjects Release \citep{Szalkai2016d,Kerepesi2016b,Kerepesi2015b}, later we have published 1064 human connectomes, each in five resolutions \citep{Varga2020}, which were based on the 1200 Subjects Release of the Human Connectome Project. For each set and each resolution, the edges of the graphs are weighted by the fiber-numbers detected between their endpoints. 

The 1064 braingraphs, described in \citep{Varga2020}, were computed by using probabilistic algorithms, and typically, any two runs on the same input yielded slightly different outputs. For increasing the robustness, and the reproducibility, we have applied an averaging and extreme-value deleting strategy, as follows:

\subsection*{Basic Averaging Strategy}

\begin{itemize}
	\item For all subjects, the tractography step of the processing, which determined the axonal fibers, connecting the ROIs of the brain, was computed 10 times;
	
	\item For each subject and each resolution (i.e., 83, 129, 234, 463, and 1015 nodes), the braingraph of the subject was computed, and ten interim weights were assigned for each edge. The ten interim weights corresponded to the number of fibers detected in the 10 tractography runs, respectively;
	
	\item Those edges, which appeared with 0 fibers in at least one of the 10 tractography runs, were deleted;
	
	\item For the remaining edges, the maximum and minimum edge-weights were deleted, and the remaining eight weights were averaged (by simple arithmetic mean). This value was assigned to the edge as its (final, non-interim) weight.
\end{itemize}

In \citep{Varga2020} the particular choice of 10 repetitions is analyzed and explained in detail.

In the present contribution, we modify the averaging strategy above and define the Newtonian blurring for braingraphs. 

We remark that the averaging process called Basic Averaging Strategy \citep{Varga2020} clearly increased the robustness of the results and decreased the variance of the graph weights due to the probabilistic processing. In the introduction of the Newtonian blurring, we also focus on robustness, and we do not intend to add any artifacts to the data: we just slightly perturb the error-correcting, averaging steps of the real, unmodified data. 

We also note that even in the augmented graphs, we have used averaging error correction for increasing robustness.

\section*{Discussion and Results} 

\subsection*{Newtonian Blurring} Here, we describe the new Newtonian Blurring method as a modification of the Basic Averaging Strategy for braingraphs. 

\begin{itemize}
	\item For all subjects, the tractography step of the processing, which determined the axonal fibers, connecting the ROIs of the brain, is computed 10 times;
	
	\item For each subject and each resolution, the braingraph of the subject is computed, and ten interim weights were assigned for each edge;
	
	\item Those edges, which appeared with 0 fibers in at least one of the 10 tractography runs, are deleted;
	
	\item For each subject and each resolution, 7 graphs from the 10 repeatedly computed graphs are chosen in every possible way (i.e., ${10\choose7}=120$ ways). Now, for each edge, the maximum and minimum edge-weights out of the 7 are deleted, and the remaining five weights are averaged (by simple arithmetic mean). This value is assigned to the edge as its (final, non-interim) weight.
\end{itemize}

This way, we augment the starting braingraph-set to a 120-times larger set. The augmented braingraphs do not contain any artificial components: they even contain strict error correction by deleting the extremal edge weights and averaging the remaining five weights.

\subsection*{On the deviation of the augmented graphs}

The following is a natural question: How diverse are the augmented graphs, compared to the starting ones? Clearly, one needs certain diversity in the augmented set for the applications in artificial intelligence since simply repeating the starting graph several times is useless. However, as in the Gaussian blurring, where the blurred images are very similar to the original ones, here we can also find a similar situation: the Newtonian-blurred graphs are also similar to the starting one. 

For a distance measure, we apply Jaccard-distance for the edges $E(G)$ of graphs $G$:

$$J(G_1,G_2)={{|(E(G_1)\cup E(G_2))-(E(G_1)\cap E(G_2))|}\over |E(G_1)\cup E(G_2)|}$$

The Jaccard-distance of any two graphs is a non-negative number between 0 and 1; it describes the fraction of edges in which the graphs differ.

Figure 1 depicts the distribution of the Jaccard-distances of all the pairs formed from the 120 augmented graphs of the two closest (in Jaccard-distance) graphs from 1053, the graphs No. 101915 and 654350. 

From the 120+120=240 graphs we can form ${240\choose 2}=28680$ pairs. The pairs are partitioned in three classes in Figure 1:

\begin{itemize}
	\item Red class: both members of the pair belong to subject 101915;
	
	\item Blue class: both members of the pair belong to subject 654350;
	
	\item Green class: one member belong to subject 101915, the other to 654350.
\end{itemize}

On the x-axis, the Jaccard distance is given; on the y-axis, the count of the pairs of graphs with the given Jaccard-distance (it is a histogram). 

In the figure, one can observe that the red, blue, and green classes form Gaussian distributions. Additionally, the expectation of the fraction of the different edges within the 120 graphs of the same origins are a little larger than 2\%, while the expectation of the Jaccard-distance in the green class is more than 0.14, i.e., at least 14\% of the edges differ.

Therefore, typically, more than 2\% of the edges differ in the graphs of the same subject, and in more than 14\% in graphs of different subjects.

\begin{figure}[H]
	\begin{center}
		\includegraphics[width=13cm]{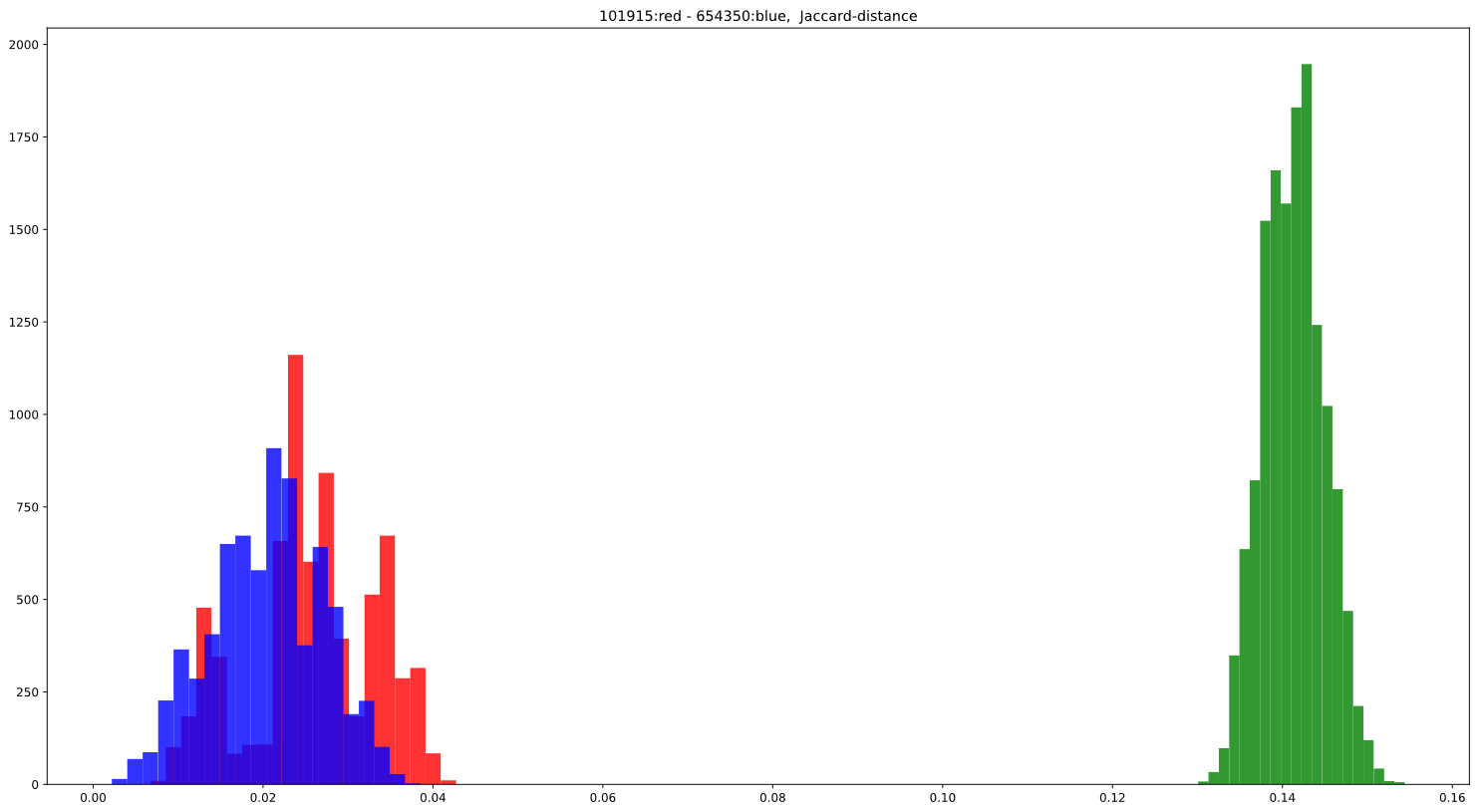}
		\caption{The distribution of the Jaccard-distances of all the pairs formed from the 120 augmented graphs of the two closest (measured in Jaccard-distance) graphs from the 1053 ones: the graphs No. 101915 and 654350. From the 120+120=240 augmented graphs, we can form 28680 pairs. The pairs are partitioned into three classes: Red class: both members of the pair belong to subject 101915; Blue class: both members of the pair belong to subject 654350; Green class: one member belongs to subject 101915, the other to 654350. On the x-axis, the Jaccard distance is given; on the y-axis, the count of the pairs of graphs with the given Jaccard-distance is shown (it is a histogram). }
	\end{center}
\end{figure}

\section*{Conclusions}

We have introduced a new augmenting method for non-image data, called Newtonian blurring. With the new method, we have prepared 631,800 augmented braingraphs for artificial intelligence applications (training and testing) and made the set publicly available at the \url{https://braingraph.org/cms/download-pit-group-connectomes/}. One important feature of the Newtonian blurring that it would not introduce any artifacts, any artificial perturbation into the data, and, still, it is capable of considerable augmenting for AI applications.
	
\section*{Data availability} The data source of this work is published at the Human Connectome Project's website at \url{http://www.humanconnectome.org} \cite{McNab2013}. The parcellation data, containing the anatomically labeled ROIs, is listed in the CMTK nypipe GitHub repository \url{https://github.com/LTS5/cmp_nipype/blob/master/cmtklib/data/parcellation/lausanne2008/ParcellationLausanne2008.xls}.

The braingraphs, computed by us, can be accessed at the  \url{http://braingraph.org/cms/download-pit-group-connectomes/} site. For each resolution, a compressed archive is available, containing 126,360 graphs. Technical remark: uncompressing an archive containing that number of files takes several minutes, even on fast systems.

\section*{Acknowledgments}
Data were provided in part by the Human Connectome Project, WU-Minn Consortium (Principal Investigators: David Van Essen and Kamil Ugurbil; 1U54MH091657) funded by the 16 NIH Institutes and Centers that support the NIH Blueprint for Neuroscience Research; and by the McDonnell Center for Systems Neuroscience at Washington University. LK, ES, BV and VG were partially supported by the VEKOP-2.3.2-16-2017-00014 program, supported by the European Union and the State of Hungary, co-financed by the European Regional Development Fund, LK, ES and VG by the  NKFI-127909
grants of the National Research, Development and Innovation Office of Hungary. LK, ES and VG were supported in part by the EFOP-3.6.3-VEKOP-16-2017-00002 grant, supported by the European Union, co-financed by the European Social Fund.
\bigskip 

\noindent Conflict of Interest: The authors declare no conflicts of interest.

\section*{Author Contribution} LK and ES invented the novel Newtonian blurring method for braingraphs, BV invented the Basic Averaging Strategy, constructed the image processing system, computed the augmented braingraphs, analyzed the deviation of the braingraphs and prepared the figures, VG has secured funding, analyzed data, and wrote the paper.

%\section*{References}

%\bibliography{v:/vince/CIKKEK/medl}
%\bibliographystyle{unsrtnat}

\end{document}